\documentclass[fleqn,10pt]{wlscirep}

\usepackage{amsmath}
\usepackage{graphicx}
\usepackage{dcolumn}
\usepackage{bm}
\usepackage{float}
\usepackage{todonotes}

\title{Spin-orbit interaction in a high-power laser irradiated micro-scale plasma waveguide}

\author[1,*]{Longqing Yi}
\author[1]{Ke Hu}
\author[1]{T\"unde F\"ul\"op}
\affil[1]{Department of Physics, Chalmers University of Technology, 41296 Gothenburg, Sweden}

\affil[*]{corresponding author: longqing@chalmers.se}

\begin{abstract}

\textbf{Light carries angular momentum as spin and orbital components. The spin-orbit interaction (SOI) of light refers to phenomena in which the spin (left or right circular polarisation) affects the spatial degrees of freedom. Recently, interest in SOI has surged \cite{Bliokh2015a}, as it provides physical insight into the behaviour of polarised light at subwavelength scales \cite{Onoda2004,Hosten2008,Bliokh2008,Bliokh2015b}, and allows for spin-controlled manipulation of light \cite{Dorney2019}. Most studies are performed with low intensity, leaving the role SOI plays in the relativistic laser-plasma interaction, characterised by nonlinearity \cite{Mourou2006}, less understood. Here, using 3D particle-in-cell simulations, we report SOI effects in this unprecedented regime. Specifically, a circularly polarised Gaussian laser irradiating a micro-scale plasma waveguide drives a spin-controlled chiral surface wave, allowing the reflected harmonic photons \cite{Teubner2009} to gain orbital angular momentum (OAM). These effects produce intense optical vortices in the extreme ultraviolet regime -- an area of significant fundamental and applied physics potential.}
\end{abstract}

\begin{document}

\flushbottom
\maketitle

\thispagestyle{empty}   

With the introduction of the chirped pulse amplification technique \cite{CPA},  relativistically intense lasers became widely available ($a_0 \equiv eE_l/m_{\rm{e}}c\omega_0 > 1$, where $E_l$ is the laser electric field, $e$  the elementary charge, $m_{\rm{e}}$  the electron mass, $c$  the vacuum light speed and $\omega_0$ the laser frequency). A laser pulse at such intensity interacts with optical media that are necessarily plasma. To study wavelength-scale effects, such as SOI, at relativistic intensities, high-contrast pulses are required to avoid amplified spontaneous emission destroying the target. Owing to development in laser cleaning techniques \cite{Thaury2007}, phenomena related to spin- or orbit-degrees of freedom of light are currently being intensively studied by the high-power laser plasma community \cite{Shi2014,Zhang2015,Zhang2016,Vieira2016,Leblanc2017,Denoeud2017}. In particular, the SOI in the high harmonics generation (HHG) process is of great interest \cite{Zhang2015,Zurch2012,Garcia2013,Dorney2019}, not only helping to resolve fundamental questions such as conservation of angular momentum in highly nonlinear laser-matter interaction, but enabling many applications from quantum optics to material characterisation \cite{Torrres2011}.

In this letter, we report the first numerical study of SOI effects in laser-plasma interaction involving a micro-scale plasma waveguide (MPW). The interaction of a laser with such targets allows light structure manipulation, leading to applications in electron acceleration and radiation generation \cite{Yi2016,Yi2019}, as recently demonstrated in experiments \cite{Snyder2019}. We show the spin (polarisation) of light provides an additional degree of freedom to control the laser-solid interaction in a MPW. This can be harnessed to reach new capabilities in producing intense extreme ultraviolet optical vortices and electron beams that carry OAM.\\


Our main results are summarised in Fig.~1: a high-power laser entering a MPW modifies the inner surface of the waveguide due to the strong electric field. This deforms the MPW and induces wavelength-scale fluctuations, as shown in Fig.~1(b): the light-green (low-density) ``ridges" and dark-green (high-density) ``valleys" are periodically distributed, forming a surface wave co-propagating with the laser. Importantly, the surface wave is chiral for a circularly polarised (CP) laser, and the screw direction is controlled by the polarisation state. Such fluctuations in turn modify light propagation, and the reflected light deviates from the plane of incidence by a small angle, analogous to the angular Imbert-Fedorov shift \cite{Bliokh2013}. This allows the reflected photons to gain intrinsic OAM due to the cylindrical geometry -- see top-right inset in Fig.~1(b).

\begin{figure}[!t]
\centering
\includegraphics[width=15.5cm]{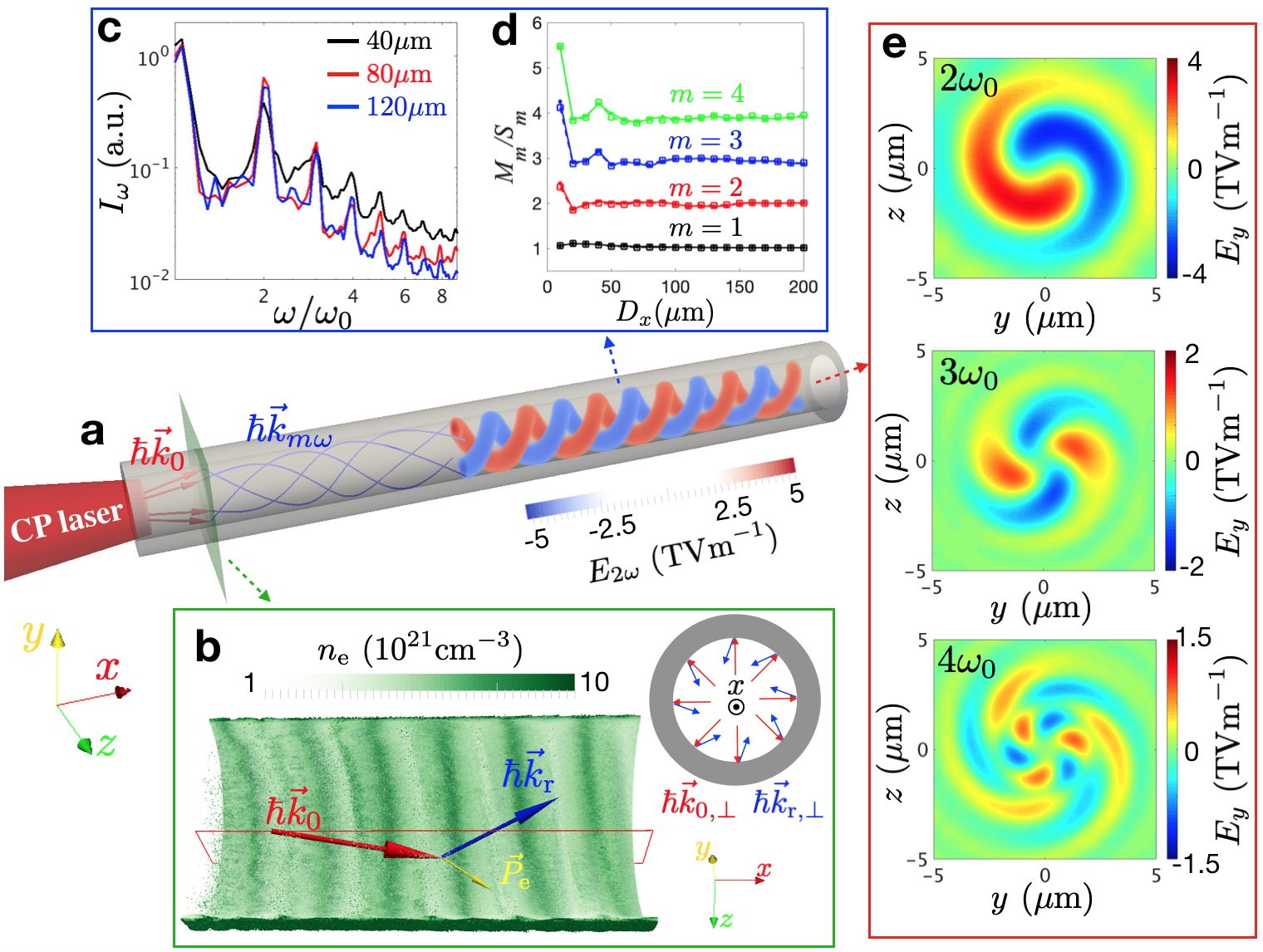}
\caption{\textbf{Generation of high-order harmonics of optical vortex in a MPW.} (a) Schematic proposed setup: a circularly polarised laser enters a MPW (grey cylinder, centre axis is the $x$-axis) from the left ($-x$). As the photons bounce between the walls, they gain spin-dependent OAM (blue curves), and the generated high-order harmonics light acquires LG-like structure (red-blue vortex, colour represents the electric field $E_y$). (b) The white-green colour shows the density fluctuation on the bottom-half of the inner surface ($y<0$) driven by a RCP laser. Due to the modulation on the surface, reflected light (blue arrow) deviates from the incident plane (red rectangle) and gains OAM,  illustrated by the inset, top-right corner of (b) (view from the $+x$ direction). (c) The energy spectrum of the electromagnetic wave in the waveguide at propagation distance $D_x=40$ $\rm{\mu m}$ (black), 80 $\rm{\mu m}$ (red), and 120 $\rm{\mu m}$ (blue). (d) Ratio $M_m/S_m$ plotted as a function of propagation distance for harmonics of order $m = 1\sim4$. Solid dots and open squares correspond to RCP and LCP drive lasers, respectively. (e) The $E_y$ field distribution for the second-, third-, and fourth-order harmonics at cross-section $x = 220{\rm{\mu m}}$. These are LG modes with topological charge $l = 1,2,3$, respectively.}
\label{fig:1}
\end{figure}


The deformation of MPW is caused by collective surface electron oscillations driven by the laser.  As the laser intensity is high ($I_0 = 2\times10^{20} {\rm W {cm^{-2}}}$), these electrons are oscillating at relativistic velocities. The photons reflected at such a surface naturally experience a relativistic Doppler shift.
Therefore HHG is produced by the relativistically oscillating mirror mechanism \cite{Bulanov1994}, which gives rise to a universal power-law spectrum in a single reflection on a sharp plasma-vacuum surface \cite{Baeva2006}. However, as the light bounces between the MPW walls, multiple reflections occur and the specific structure of the spectrum evolves with propagation distance, shown in Fig.~1(c). The harmonics are generated mostly in the first few tens of microns, then their intensity decreases due to dephasing (discussed later). 


To study the SOI effects, we introduce a simple ``shaken waveguide" model. As the first-order approximation, we assume the cross-section of the MPW remains circular but its centre is shifted in the negative direction of the radial electric field of the fundamental harmonic ($\omega_0$), which can be obtained from plasma waveguide theory (zeroth-order approximation, see Supplementary Material) as:
\begin{equation} 
E_{r}(x,r,\phi) \approx E_0{\rm J}_0(k_{\rm T}r)\exp(i\sigma\phi)\exp(ik_xx-i\omega_0t),
\label{eq1}
\end{equation}
where the radial coordinate $r = \sqrt{y^2+z^2}$, and the azimuthal coordinate $\phi$ is measured counter-clockwise with respect to the $y$-axis in the $y$-$z$ plane. $E_0$ is the amplitude, ${\rm{J_0}}(x)$ is the Bessel function of the first kind, $k_x$ and $k_{\rm{T}}$ are the longitudinal and transverse wavenumber components, satisfying $k_x^2 + k_{\rm{T}}^2 = k_0^2$, with $k_0$ the wavenumber in vacuum. In a waveguide with radius $r_0\approx2.8 \rm {\mu m}$, the transverse wavenumber is given by the the smallest root of the eigenvalue equation: $k_{\rm{T}} = 2.35/r_0$. We have used $k_{\rm T}\ll k_0$ in Eq.~(1). Notably, the SOI effects stem from the $\exp(i\sigma\phi)$ term, where $\sigma = \pm1$ is controlled by the polarisation state of the drive laser: $\sigma = +1$ for right-handed circular polarisation (RCP) and $\sigma = -1$ for left-handed (LCP).

We now set the phase of Eq.~(1) to be the phase of the waveguide centre shift: $\delta r = -\delta r_0\exp(i\sigma\phi)\exp(ik_xx-i\omega_0t)$, where $\delta r_0\approx a_0'c/\omega_0 \ll r_0$ is the amplitude, and $a_0'$ is the normalised electric field near the inner boundary. Substituting this running periodic displacement of the waveguide centre back into Eq.~(1) with $r \rightarrow r-\delta r$ (similarly to the expression for the azimuthal component $E_\phi$), and using a Taylor expansion, the transverse electric field can be written as a sum of harmonics:
\begin{equation} 
\mathbf{E}_{\perp}(x,r,\phi) \approx (\mathbf{y}+i\sigma\mathbf{z})\sum_{m=1}^{\infty}(k_{\rm T}\delta r_0)^{m-1}E_{0}\frac{{\rm J}_0^{(m-1)}(k_{\rm T}r)}{(m-1)!}\exp[i(m-1)\sigma\phi]\exp(imk_xx-im\omega_0t),
\label{eq2}
\end{equation}
where $m$ denotes the harmonic order.

The SOI effect is transparent in Eq.~(2): the spin angular momentum (SAM) of the incident CP light is transformed into intrinsic OAM of harmonics, giving rise to Laguerre-Gaussian(LG)-like modes \cite{Allen1992} with well-defined topological number $l = (m-1)\sigma$. The polarisation state (SAM of photons) does not change in the HHG process.

Each photon, characterised by $\sigma$ and $l$, carries SAM $\mathbf{S^{\rm ph}} = \sigma\hbar\mathbf{x}$ and OAM $\mathbf{L^{\rm ph}} = l\hbar\mathbf{x}$ parallel to the propagation direction, and the total angular momentum (TAM) of a photon is $\mathbf{M}^{\rm ph} = \mathbf{S^{\rm ph}}+\mathbf{L^{\rm ph}}$. When $m$ photons at the fundamental frequency are transformed into one photon of $m$th-order harmonic, their SAM ($m\sigma\hbar$) go to the same harmonic order, in the form of $\sigma\hbar$ SAM plus $(m-1)\sigma\hbar$ OAM. Thus, the conservation of TAM and energy are guaranteed. 

This result is demonstrated by particle-in-cell (PIC) simulations. In Fig.~1(d), the ratio $M_m/S_m$ (for $m =1\sim4$) is plotted against propagation distance, where $M_m = \epsilon_0[\int\mathbf{r}\times(\mathbf{E}_m\times\mathbf{B}_m){\rm  d}\mathbf{r}]_{x}$, and $S_m = \sigma W_m/m\omega_0$ (with $W_m$ denoting the field energy) are respectively sums of longitudinal components of TAM and SAM of all the photons of the $m$th harmonic. After the initial phase (discussed later), a simple relation $M_m/S_m = m$ is obtained. From a quantum optics point of view, with the polarisation state being unchanged, the relation $l = (m-1)\sigma$ is confirmed. This is further verified by the transverse profile of $E_y$ fields shown in Fig.~1(e).\\

\begin{figure}[!b]
\centering
\includegraphics[width=15.5cm]{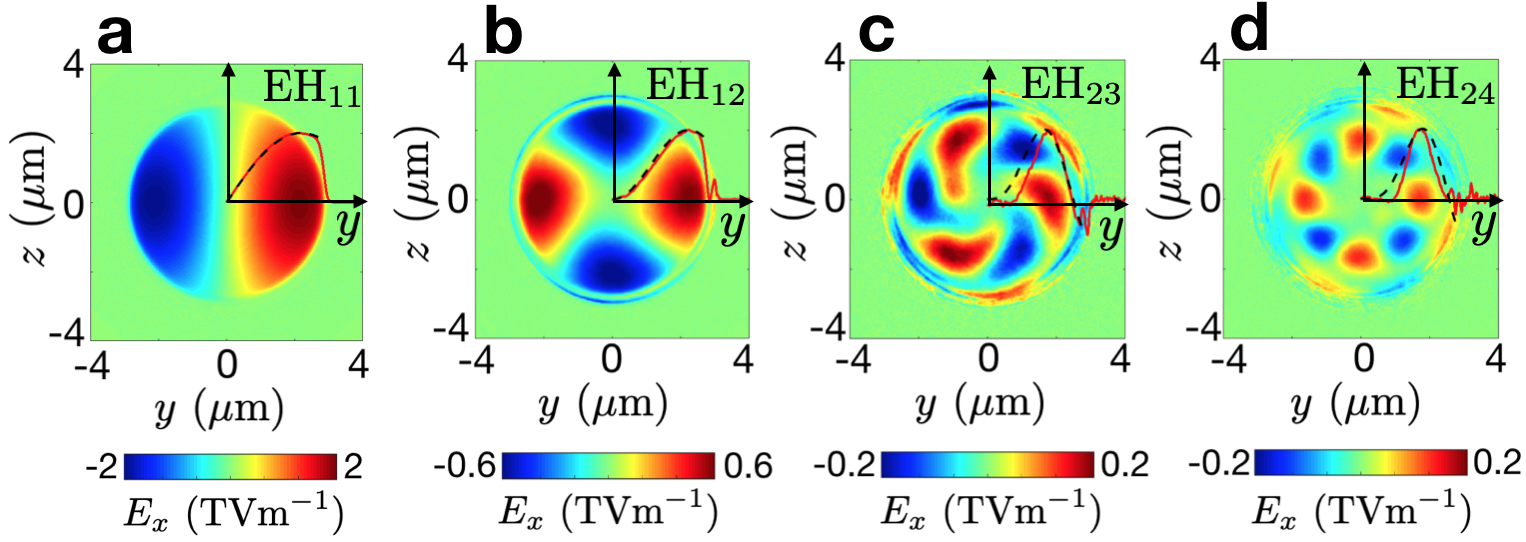}
\caption{\textbf{Waveguide modes of the harmonics.} Longitudinal electric field $E_x$ of (a) first- (b) second- (c) third- and (d) fourth-order harmonics inside MPW at propagation distance $D_x = 200{\rm{\mu m}}$. The red solid lines are the 1D field profiles  along the $y$-axis ($y>0$ side), and the black dashed lines are fits of the waveguide mode  \rm{EH}$_{11}$ ($\omega_0$), \rm{EH}$_{12}$ ($2\omega_0$), \rm{EH}$_{23}$ ($3\omega_0$), and \rm{EH}$_{24}$ ($4\omega_0$), respectively.}
\label{fig:2}
\end{figure} 


Note, SOI is forbidden for the fundamental harmonic $l(m=1) = 0$, so the nonlinear effects (and therefore, use of a high-power laser) are essential for the proposed scheme. According to Eq.~(2), this indicates $k_{\rm T}\delta r_0 \approx a_0'k_{\rm T}/k_0>1$, which corresponds to a threshold laser intensity around $I_{\rm th}\sim 10^{20}$ Wcm$^{-2}$ for a waveguide with radius of a few microns (typically $k_{\rm T}/k_0\sim0.1$). 

Moreover, since Eq.~(2) does not automatically fulfil the boundary conditions, the generated HHG beams will be converted into waveguide modes, which usually propagate at different velocities from the drive laser. The harmonics produced at different longitudinal positions thus not add resonantly, and the dephasing, caused by the mismatch between the velocities of harmonics and the driver, results in a dynamic spectrum as shown in Fig.~1(c). The intensities of harmonics increase within dephasing length, after which they decay. Therefore the length of MPW should be carefully chosen to prevent the harmonics of interest from fading out.

The dephasing effect has a profound impact on the waveguide modes of the harmonics. In general, a variety of modes (EH$_{pq}$) are excited at each harmonic order, where $q$ is the number of periods in the $E_x$ field within $\phi = 0\sim2\pi$, related to the topological charge $l$ by $q = l+1$; and $p$ is the $p$th-smallest root of the eigenvalue equation (see Supplemental Material). At the fundamental harmonic (Fig.~2(a)), the lowest-order mode dominates, which drives the ``shaken waveguide". However, for the HHG beams, due to the dephasing effect, modes with velocities closer to the fundamental harmonic are more likely to survive. This gives rise to the high-order waveguide modes shown in Fig.~2(b-d). In particular, the lowest-order mode ($p=1$) at the third- and fourth-harmonics have almost faded out. \\


The nonlinear nature of high-power laser-plasma interaction allows phenomena absent in the low-intensity regime to come into play. As shown in Fig.~3, some electrons are ejected into the vacuum core of the MPW as a result of a wave breaking effect, i.e.~when the surface electrons oscillate in the surface wave, some move too quickly toward the axis and do not return to the wall. These electrons are accelerated and co-propagate with the laser \cite{Yi2016}. For a CP laser, the accelerated electrons will also carry OAM that depends on the polarisation state, leading to the SOI of light and matter.

\begin{figure}[!b]
\centering
\includegraphics[width=15.5cm]{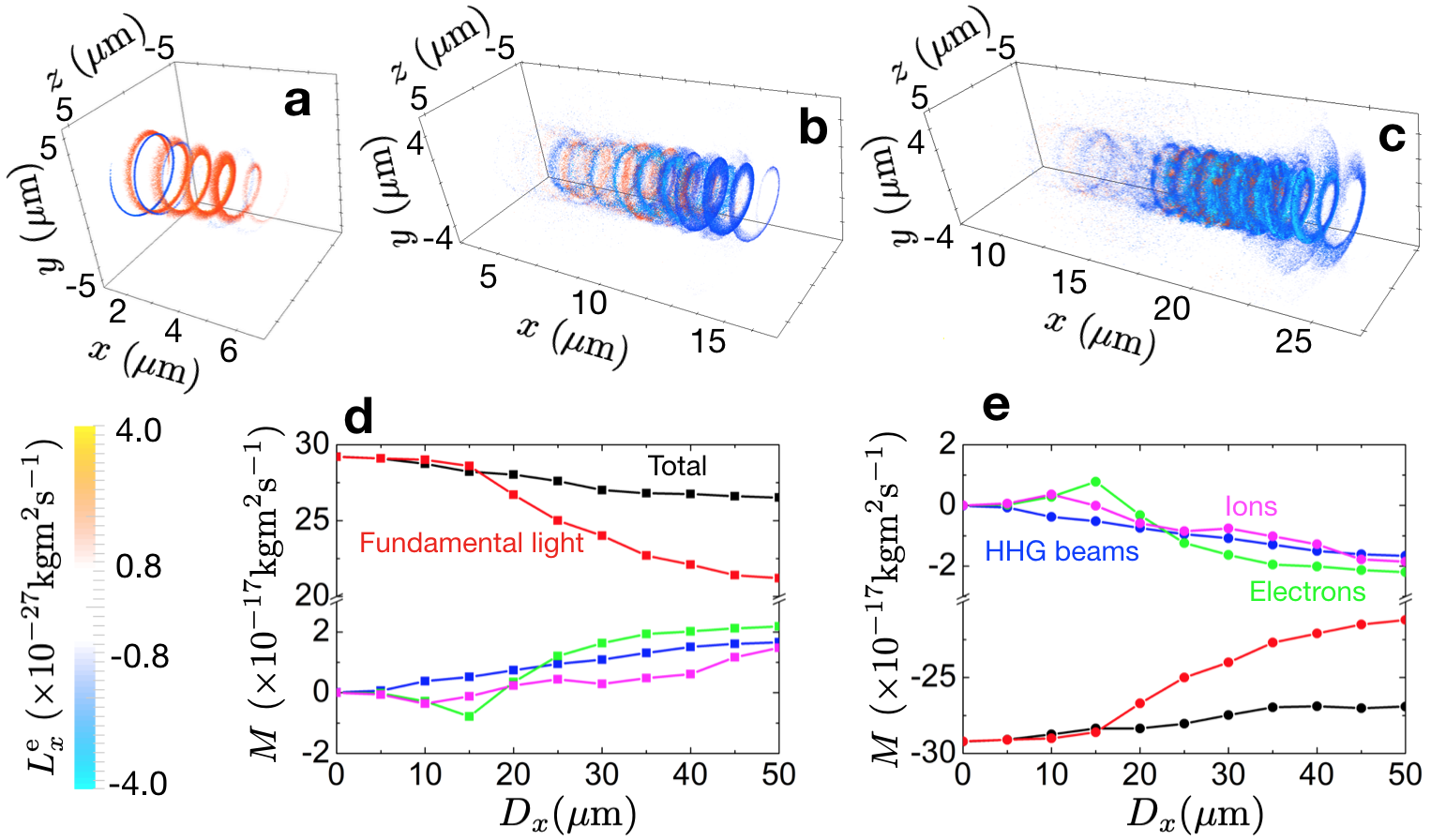}
\caption{\textbf{Generation of helical electron beams.} Relativistic electron beam (with longitudinal momentum $P_x/m_{\rm{e}}c>2$) carrying OAM produced by a LCP laser interaction with MPW, shown at the propagation distance (a) 5 ${\rm \mu m}$ (b) 15 ${\mu m}$, and (c) 25 ${\mu m}$. Colour shows the longitudinal OAM of each electron. The evolution of TAM carried by components in the simulation with propagation distance is shown for RCP (d) and LCP (e) laser. The fundamental light ($m=1$), high-order harmonics ($m >1$), electrons, and ions are colour-coded with red, blue, green and magenta, respectively. The black curve shows the sum of all components.}
\label{fig:1}
\end{figure}

The 3D structure of the accelerated electron beam is shown in Fig.~3(a-c) for a LCP incident laser: a helical structure is formed,  the colour representing the longitudinal OAM of individual electrons $L_x^{\rm{e}} = yP_z - zP_y$, where $P_j$ is the linear electron momenta in the $j$-direction ($j$ being $x$, $y$, or $z$). The evolution of electron OAM is more sophisticated. As indicated by Fig.~1(b), electrons initially experience a recoil force due to the deviation of the reflected light, giving rise to an inverse OAM with respect to the SAM of the laser (see Fig.~3(a)).

Generating the electron beam takes energy from the light, resulting in a reduction of photon number. As these electrons carry an opposite sign of angular momentum with respect to the photons absorbed, the remaining light must carry excess OAM to conserve the TAM of light and matter. This is responsible for the deviation from Eq.~(2) in the early period shown in Fig.~1(d). However, this excess OAM is soon transferred to the electrons by subsequent interaction causing them to orbit the beam axis, with the same vorticity as the light \cite{Chavez2003}. Thus the electron OAM is reversed again (see Fig.~3(b-c)), a strong indication that the light inside the MPW has gained intrinsic OAM.

Finally TAM conservation is illustrated by the black curves in Fig.~3(d-e) for the RCP and LCP laser, respectively. A slight drop is due to loss from the simulation box of the fraction of light reflected at the MPW entrance. The reversal of the particle OAM is seen in both cases at the early phase. Moreover, the simulations indicate that within 50-${\rm \mu m}$ propagation, about $20\%$ of the laser's SAM is transformed into OAM of high-order harmonics and particles, resulting in a relativistically rotating bunch. Using $L_x^{\rm e} = \gamma_{\rm e} m_{\rm e}\omega_{\rm e} r^2$, the angular velocity of electrons in Fig.~3(c) can be estimated to be $\omega_{\rm e} \sim10^{12}~{\rm rad~s^{-1}}$ (assuming Lorentz factor $\gamma_{\rm e} \sim 100$). This is of interest in relativistic laboratory astrophysics, such as the relativistic rotator \cite{Bulanov2009}.\\

In conclusion, we demonstrate the generation of circularly polarised high harmonic extreme ultraviolet optical vortices with topological charge $(m-1)\sigma$ ($m$ being the harmonic order and $\sigma = \pm1$ denotes the polarisation state) originating from SOI when a high-power CP laser propagates in a MPW. In this process, a helical electron beam is produced that rotates about the laser axis with relativistic velocities. We demonstrate SAM-OAM conversion and  conservation of TAM in the nonlinear HHG process. This can be a powerful tool for controlling the laser-matter interaction and producing HHG beams with a designed OAM. This opens a new and promising perspective in ultrafast science and relativistic laser-plasma interaction.\\

\section*{Methods}
\textbf{Laser-plasma parameters.} We simulate a circularly-polarised high-power laser beam entering the simulation box from the left ($-x$) boundary, propagating to the right. The laser field is $\mathbf{E}_l = (\mathbf{y}+i\sigma\mathbf{z})E_{l0} \exp(-r^2/w_0^2) \sin^2(\pi t/\tau_0)\exp(ik_0x-i\omega_0 t)$, where $0<t<\tau_0 = 54$ fs, $E_{l0} = 32$ TV m$^{-1}$ is the laser amplitude (corresponding to a normalised amplitude $a_0 = 10$ and intensity $I_0 = 2\times10^{20}$ W~cm$^{-2}$), spot size $w_0 = 2.5 {\rm \mu m}$, frequency $\omega_0 = k_0c$, wavenumber $k_0 = 2\pi/\lambda_0$ with $\lambda_0 = 1 {\rm \mu m}$  the laser wavelength. The laser polarisation state is controlled by $\sigma = +1$ for RCP and $-1$ for LCP. 

The target considered in this work is a small channel of a few microns in glass or plastic material, being widely available as a standard micro-channel plate used for X-ray detection \cite{Gys2015}. In the simulation, the MPW (assumed plastic(CH)) is modelled by a pre-ionised cylindrical plasma with electron density $n_0 = 300n_{\rm{c}}$, with $n_{\rm{c}} = m_{\rm{e}}\omega_0^2/4\pi e^2 = 1.1\times 10^{21}$ cm$^{-3}$ the critical density (corresponding to a mass density $\sim 1$ g~cm$^{-3}$). The inner and outer radii of the MPW are $r_{\rm{i}} = 4 {\rm \mu m}$ and $r_{\rm{o}} = 5 {\rm \mu m}$, respectively, and the longitudinal MPW length is $200 {\rm \mu m}$. Representing heating by the laser pre-pulse, the inner surface ($r<r_{\rm{i}}$) is assumed to have a density gradient $n(r) = n_0\exp(-(r-r_{\rm{i}})^2/h^2)$, and scale length $h = 0.5 {\rm \mu m}$. The effective radius of MPW ($r_0$) for calculating the roots of eigenvalue equation is where the MPW wall becomes overdense, namely $n(r_0) = 1n_{\rm c}$.

For the simulation results presented in Fig.~3(d-e), a reduced density $n_0 = 55 n_{\rm{c}}$ is used to improve computational efficiency, as a larger simulation box is used (motivated below). The other parameters are kept the same. The change in maximum density makes little difference in the results as the laser can hardly penetrate  to the area above 55 $n_{\rm{c}}$.\\
 
{\noindent}\textbf{PIC simulation.} The 3D PIC simulations presented in this work were conducted with the {\sc epoch} code \cite{Arber2015}. For the primary simulation results presented in Fig.~1, Fig.~2, and Fig.~3(a-c), the dimensions of the simulation box are $x\times y\times z = 25 {\rm \mu m}\times 11 {\rm \mu m} \times 11 {\rm \mu m}$, sampled by $2000\times440\times440$ cells with five macro particles for electrons, two for C$^{6+}$ and two for H$^{+}$ ions. A moving window is used to improve computational efficiency, which follows the laser propagating in the MPW at light velocity along the $x$-axis. 

The simulation for Fig.~3(d-e) used a larger simulation box $x\times y\times z = 70 {\rm \mu m}\times 11 {\rm \mu m} \times 11 {\rm \mu m}$, sampled by $2100\times220\times220$ cells, with the number of macro particles per cell the same. The moving window is turned off to avoid particles carrying angular momentum to stream out of the simulation box. In both cases, a high-order particle shape function (fifth order particle weighting) is applied to reduce numerical self-heating instabilities (see Sec. 5.1 of Arber et al.~\cite {Arber2015} for details).

As the numerical dispersion relation can be vital for the coherence condition for HHG, the algorithm developed by Cowan et al.~\cite{Cowan2013} is used to minimise the numerical dispersion, the results are presented in the figures. We also use the NDFX scheme proposed by Pukhov \cite{Pukhov1999}, which aims to improve the numerical dispersion relation, to cross-check these results (not shown). In both schemes, the numerical convergence has been confirmed by comparing the HHG spectra for the simulations with different resolutions.

\subsection*{Data availability.}
The data that support the findings of this study are available from the corresponding author upon request.

\section*{Acknowledgements}
The authors are grateful to A. Pukhov, S.L. Newton, and I. Pusztai for fruitful discussions. This work is supported by the Olle Engqvist Foundation, the Knut and Alice Wallenberg Foundation and the European Research Council (ERC-2014-CoG grant 647121). Simulations were performed on resources at Chalmers Centre for Computational Science and Engineering (C3SE) provided by the Swedish National Infrastructure for Computing (SNIC).

\section*{Author contributions statement}
 L.Q.Y. conceived the idea and designed the setup. L.Q.Y. and K.H. conducted the simulations and analysed the results, under supervision of T.F.. L.Q.Y. and T.F. wrote the paper.

\section*{Additional information}
Competing interests: The authors declare no competing interests.

\end{document}